\documentclass[final,3p,times]{elsarticle}

\usepackage{amssymb}
\usepackage{amsmath}

\usepackage{calc}
\usepackage{array}
\usepackage{booktabs}
\usepackage{makecell}
\usepackage{multirow}
\usepackage{subfigure}
\usepackage{longtable}
\usepackage{url}
\usepackage[ruled,vlined,linesnumbered,commentsnumbered]{algorithm2e}
\usepackage{hyperref}
\usepackage{color}
\usepackage{tcolorbox}
\usepackage{xcolor,colortbl}

\newcommand{\chen}[1]{\textcolor{black}{#1}}

\journal{arXiv}

\begin{document}

\begin{frontmatter}



\title{Inference performance evaluation for LLMs on edge devices with a novel benchmarking framework and metric}



\author[label1]{Hao Chen} 
\author[label1]{Cong Tian}
\author[label1]{Zixuan He}
\author[label1]{Bin Yu}
\author[label2]{Yepang Liu}
\author[label3]{Jialun Cao}

\affiliation[label1]{organization={Xidian University},
            addressline={No.2 South Taibai Road}, 
            city={Xi'an},
            postcode={710071}, 
            state={Shaanxi},
            country={China}}

\affiliation[label2]{organization={Southern University of Science and Technology},
            addressline={1088 Xueyuan Avenue}, 
            city={Shenzhen},
            postcode={518055}, 
            state={Guangdong},
            country={China}}

\affiliation[label3]{organization={The Hong Kong University of Science and Technology},
            addressline={Clear Water Bay, Kowloon}, 
            city={Hong Kong},
            country={China}}

\begin{abstract}
With the significant success achieved by large language models (LLMs) like LLaMA, edge computing-based LLM inference services for mobile and PC are in high demand for data privacy. However, different edge platforms have different hardware characteristics and the large demand for memory capacity and bandwidth makes it very challenging to deploy and benchmark LLMs on edge devices. In this paper, we introduce a benchmarking tool named ELIB (edge LLM inference benchmarking) to evaluate LLM inference performance of different edge platforms, and propose a novel metric named MBU to indicate the percentage of the theoretically efficient use of available memory bandwidth for a specific model running on edge hardware to optimize memory usage. We deploy ELIB on three edge platforms and benchmark using five quantized models to optimize MBU in combination with other metrics such as FLOPS, throughput, latency and accuracy. And we analyze the results to derive the key factors, constraints, unpredictability in optimizing MBU that can guide deploying LLMs on more edge platforms.
\end{abstract}




\end{frontmatter}



\section{Introduction}

Transformer \cite{Vaswani2017AIA} based large language models (LLMs), like ChatGPT \cite{Brown2020LMA,Ouyang2022TLM}, ChatGLM \cite{Du2021GGL,Zeng2022G1A} LLaMA \cite{Touvron2023LOA}, have achieved considerable success in rencent years, demonstrating outstanding performance in chatbots \cite{Shahriar2023LHA}, photo generation robots \cite{Oppenlaender2022TCO}, etc. As LLMs grow larger and more complex, LLMs are typically trained and deployed on cloud servers. However, inference LLMs on cloud servers suffer from two main issues, i.e., \textbf{data privacy and security issue}. Specifically, on the one hand, in LLM application scenarios such as chatbots and local knowledge bases, certain data are confidential and cannot be uploaded to remote cloud servers. On the other hand, \textbf{latency and robustness issues}: in application scenarios like self-driving cars, remote access to cloud servers is not permitted on safety-critical systems due to bandwidth and availability. Therefore, edge computing-based LLM inference can provide a localized solution for certain application scenarios

However, deploying a LLM on an existing edge computing platform presents numerous significant challenges. First, edge computing systems \cite{Zhang2020DLI,Barbera2013TOO} encompass a wide range of platforms and are often restricted by physical limitations such as power consumption. For example, high-end Nvidia GPUs with more than hundreds of watts \cite{Jo2020BGA} power consumption may not be suitable for many limited power scenarios. Therefore, various edge deep learning platforms for PC, mobile, and IoT \cite{Ma2019ASO,Hu2016QTI} are designed to speed up the LLM inference in different edge environments.
Second, the emerging edge computing-based platforms feature diverse underlying hardware characteristics and employ various software tools for edge-based LLM implementation and adoption. The model quantization methods, software configurations, and hardware specifications diversity make it difficult to compare the performance of different platforms directly and efficiently \cite{Kong2022}. Additionally, a standardized evaluation metric for LLM inference should be defined and computed consistently across different edge devices to accurately describe the overall inference evaluation result. The state-of-the-art evaluation methods \cite{Varghese2021ASO,Bianco2018BAO,Coleman2017DAE} primarily focus on conventional performance metrics competition, which are insufficient for comprehensive evaluation. Third, model quantization is a method to decrease the numerical precision of weights and activation functions in a neural network    \cite{Kuzmin2022FQT,Micikevicius2022FFF,Rodriguez2018LNP}, reducing the computation costs of inference. Furthermore, the performance of a quantized model on an edge computing platform is closely tied to the quantization method, software configuration, and hardware specification. The optimal combination of quantization method, software configuration, and hardware specification is crucial for achieving the best trade-off between accuracy and performance in edge computing-based LLM inference.

Indeed, only knowing the best accuracy or latency for a specific LLM on an edge computing platform is insufficient for insightful information. Practically, the challenge lies in enhancing an LLM on a particular edge computing platform to achieve better accuracy without sacrificing performance. Therefore, a systematic method is required to roughly evaluate the performance limits of edge platforms under complex conditions.

In this paper, we propose a new evaluation method, ELIB, to evaluate the overall performance of different edge deep learning platforms based on widely-used LLM. ELIB will abstract and simplify the complicated deployment process of different edge platforms, especially hybrid computing. Then ELIB will evaluate overall floating-point and integer computing capability, token filled and generated rate, and inference accuracy of a certain quantitated model and device. Moreover, a novel benchmark metrics is discussed to optimize hardware memory utilization. Finally, we employ ELIB to benchmark the three most widely used edge computing platforms by different quantization models. Then we analyze the results and obtain key factors that can guide adapting the model to edge hardware.

The rest of this paper is organized as follows: Section 2 gives a short discussion about related works. Section 3 outlines the overview of the target edge devices, acceleration frameworks, LLMs, and quantitation methods used in this work for evaluation. Section 4 discusses the design overview and benchmarking metrics. Section 5 demonstrates and analyzes the benchmarking results, along with the heuristics summarized from the findings. Finally, Section 6 concludes the paper by summarizing the work.

\section{Related Work}

The edge platforms are frequently deployed in remote areas with limited access to power or even on mobile devices with very restrictive battery management \cite{Ogden2018MMD}. The performance objectives of LLMs on the edge differ from those in GPU-based cloud center environments, as platform characteristics, use cases, and benchmarking metrics typically vary significantly. The primary focus of running LLM workloads on the edge is to strike a balance between various benchmarking metrics such as throughput, latency, accuracy, power, and more.

However, the current evaluation methods \cite{Pilsung2021BME} primarily focus on conventional performance metrics competition for powerful GPU-based clouds, which are insufficient for conducting a comprehensive evaluation for edge platforms. A specific approach is required to focus on benchmarking large language models on edge computing platforms.

The current benchmarking tools for edge computing primarily focus on evaluating CNN-based computer vision tasks \cite{Baller2021DBD}. However, as LLMs are transformer-based networks with much larger model sizes compared to traditional CNN-based models, benchmarking tools like Mlperf inference benchmark \cite{Reddi2020MIB} are not suitable for transformer-based LLMs due to their distinct structure and different computing power requirements.

The community leaderboards \footnote{\url{https://huggingface.co/spaces/lmsys/chatbot-arena-leaderboard}} only list the inference accuracy of famous LLMs, but their inference performance is not widely discussed because they are all deployed on GPU-based cloud centers. Additionally, there is extensive research on benchmarking LLMs' capabilities for specific tasks such as chatbot  \cite{Banerjee2023BLP}, recommendation \cite{Liu2023LBL} and summarization \cite{Zhang2023BLL}.

Our approach differs from the method mentioned above in that we conduct a comprehensive benchmarking and comparison of LLM performance on various edge computing platforms using commonly used LLMs and performance metrics from a more practical perspective. Our goal is to provide researchers with practical heuristics to explore further the capabilities of deploying LLMs on edge computing platforms.

\section{Target Edge Devices, Acceleration Frameworks, LLMs and Quantitation Methods}

In this section, we contrast various edge device hardware architectures and associated acceleration software frameworks. We also provide an overview of the target inference frameworks and briefly outline the characteristics of different LLMs and quantization methods used for benchmarking the edge devices in this study.

\subsection{Target Edge Devices and Acceleration Frameworks}

We evaluated the three edge-devices and associated acceleration frameworks outlined in Table 1. \chen{ We selected three representative hardware platforms from IoT, mobile, and personal computer. The selection was based on a combination of market popularity, operating systems, software frameworks, hardware capabilities, and support for machine learning workloads. Our goal was to cover a diverse range of devices to ensure our findings are broadly applicable.. }

\begin{enumerate}
\item
  \emph{NanoPI}\footnote{\url{https://wiki.friendlyelec.com/wiki/index.php/NanoPC-T6}}: The NanoPI, featuring the RK3588 chip\footnote{\url{https://www.rock-chips.com/a/en/index.html}}, is an IoT development board. It utilizes a quad-core Cortex-A76 as its large core and a quad-core Cortex-A55 as its small core, delivering robust processing capabilities. With 16GB of LPDDR4x RAM boasting a bandwidth of 34GB/s, it offers substantial memory performance. The Mali-G610 Graphics Processing Unit (GPU) enables high-power graphics processing for complex applications and supports hybrid computing to accelerate numerical processing. The Nano Pi is compatible with the Ubuntu OS and acceleration frameworks such as OpenBLAS\footnote{\url{https://www.openblas.net}}, OpenCL \cite{Stone2010OAP}, and CLblast \cite{Nugteren2018CAT}.

  \begin{enumerate}
  \item
    \emph{OpenBLAS}: The OpenBLAS is optimized to leverage the capabilities of contemporary CPUs, including multi-core processors and hardware acceleration features such as SIMD (Single Instruction, Multiple Data) instructions
  \item
    \emph{OpenCL}: The OpenCL is an open standard framework for parallel programming across hybrid computing platforms, including CPUs, GPUs, and other accelerators. It provides a unified and standardized interface for writing programs that can execute on a wide range of computing devices, enabling developers to harness the power of parallel processing for a variety of computational tasks. 
  \item
    \emph{CLblast}: The CLblast is an open-source software library designed to provide efficient and optimized implementations of basic linear algebra subprograms (BLAS) for OpenCL-enabled devices, particularly GPUs. It is developed to accelerate various linear algebra operations on parallel processing devices, enhancing the performance of computations in fields such as scientific computing, machine learning, and data analysis.
  \end{enumerate}
\item
  \emph{Xiaomi Redmi Note12 Turbo}\footnote{\url{https://www.mi.com/redmi-note-12-turbo}}: This device is equipped with the Snapdragon 778 chip \footnote{\url{https://www.qualcomm.com/products/mobile/snapdragon/smartphones/snapdragon-7-series-mobile-platforms}}, featuring a CPU that combines four high-performance cores. This includes one Kryo Prime core based on the Cortex-X2 architecture clocked at 2.91 GHz, three Kryo Gold cores based on the Cortex-A710 architecture clocked at 2.49 GHz, and four energy-efficient Kryo Silver cores based on the Cortex-A510 architecture clocked at 1.8 GHz. It also boasts 16GB of LPDDR RAM with a 26GB/s bandwidth. The SoC incorporates an Adreno 725 GPU clocked at 580 MHz, delivering up to 1781.7 GFLOPS of floating-point performance in FP32 operations. It supports the Android OS and acceleration frameworks such as OpenBLAS, OpenCL, and CLblast.
\item
  \emph{MacBook Air 2022}\footnote{\url{https://www.apple.com/macbook-air-13-and-15-m2/specs/}}: This high-performance laptop is powered by the Apple Silicon M2 chip, featuring 8 cores with 4 high-performance and 4 high-efficiency. The M2 utilizes 16GB of LPDDR5 RAM in a unified memory configuration, offering a bandwidth of 50 GB/s. Additionally, it includes a ten-core GPU with up to 1280 ALUs, delivering a maximum floating-point performance of 3.6 TFLOPs. It supports MacOS and acceleration frameworks such as Apple Accelerate\footnote{\url{https://developer.apple.com/documentation/accelerate}} and Apple Metal \cite{Gebraad2023SGA}.

  \begin{enumerate}
  \item
    \emph{Apples Accelerate:} This framework is a powerful and optimized library designed to accelerate mathematical and numerical computations on Apple platforms. It provides a collection of low-level and high-performance routines for various tasks such as linear algebra, signal processing, image processing, and more.
  \item
    \emph{Apple Metal}: This is a high-performance graphics and compute API developed by Apple. It is designed to provide developers with direct access to the GPU on Apple devices. Metal offers a streamlined and efficient way to harness the full power of the GPU for rendering graphics, performing general-purpose computing (GPGPU), and accelerating various tasks.
  \end{enumerate}
\end{enumerate}

\begin{table}[!ht]
\caption{Overview of different target edge devices hardware specs and acceleration }
\begin{tabular}[]{@{}
  >{\raggedright\arraybackslash}p{(\columnwidth - 12\tabcolsep) * \real{0.1231}}
  >{\raggedright\arraybackslash}p{(\columnwidth - 12\tabcolsep) * \real{0.1236}}
  >{\raggedright\arraybackslash}p{(\columnwidth - 12\tabcolsep) * \real{0.1439}}
  >{\raggedright\arraybackslash}p{(\columnwidth - 12\tabcolsep) * \real{0.1348}}
  >{\raggedright\arraybackslash}p{(\columnwidth - 12\tabcolsep) * \real{0.1546}}
  >{\raggedright\arraybackslash}p{(\columnwidth - 12\tabcolsep) * \real{0.1237}}
  >{\raggedright\arraybackslash}p{(\columnwidth - 12\tabcolsep) * \real{0.1963}}@{}}
\toprule
Platforms & Devices & CPU & RAM & GPU & OS & Frameworks \\
\midrule
IoT 
& Nano PI 
& RK3588
8cores
2.9GHz 
& LPDDR4X
16G
2133GHz
34GB/s 
& Mali-G610 
& Ubuntu 
& OpenBLAS 

OpenCL

CLblast \\

Mobile 
& Xiaomi Redmi Node12 
& Snapdragon 778 
8 cores 
2.5GHz 
& LPDDR4
16G
1600GHz
26GB/s 
& Adreno 725 
& Android 
& OpenBLAS 

OpenCL

CLblast\\

PC 
& Apple MacBook Air 2022 
& Apple M2
8 cores
3.6GHz 
& LPDDR5
16G
3200GHz
50GB/s
& Apple CPU
& MacOS 
& Apple Accelerate
Apple Metal \\

\bottomrule
\end{tabular}
\end{table}

\begin{table}[!ht]
\caption{Summary of three native LLM inference frameworks}
\begin{tabular}[]{@{}
  >{\raggedright\arraybackslash}p{(\columnwidth - 14\tabcolsep) * \real{0.1189}}
  >{\raggedright\arraybackslash}p{(\columnwidth - 14\tabcolsep) * \real{0.1450}}
  >{\raggedright\arraybackslash}p{(\columnwidth - 14\tabcolsep) * \real{0.1450}}
  >{\raggedright\arraybackslash}p{(\columnwidth - 14\tabcolsep) * \real{0.1345}}
  >{\raggedright\arraybackslash}p{(\columnwidth - 14\tabcolsep) * \real{0.1247}}
  >{\raggedright\arraybackslash}p{(\columnwidth - 14\tabcolsep) * \real{0.1107}}
  >{\raggedright\arraybackslash}p{(\columnwidth - 14\tabcolsep) * \real{0.1060}}
  >{\raggedright\arraybackslash}p{(\columnwidth - 14\tabcolsep) * \real{0.1175}}@{}}
\toprule
 & CPU & GPU & Quantitation & Dependency & Portability & Flexibility & Complexity \\
\midrule
MLC-LLM
& 
OpenBLAS
Apple Accelerate
Intel MKL
& 
OpenCL
Vulkan
CUDA
Metal
& LUT-GEMM & TVM & Low & Low & High \\

CTranslate2 
& OpenBLAS
Apple Accelerate
Intel MKL
& CUDA OpenCL & GPTQ & No & Middle & Lox & High \\

Llama.cpp 
& 
OpenBLAS
Apple Accelerate
Intel MKL
& 
OpenCL
CLblast
Metal
& GGML & No & High & High & Low \\
\bottomrule
\end{tabular}
\end{table}

\subsection{Inference frameworks and evaluation models}
\chen{The selection of inference frameworks was driven by the specific needs of edge computing environments, prioritizing lightweight, efficient, easy-to-deploy, and scalable solutions. Common frameworks like TensorRT, ONNX Runtime, and TensorFlow Lite are powerful but heavyweight and Python-based, making them less suitable for resource-limited edge devices. In contrast, llama.cpp, a pure C++ framework, offers significant advantages in performance, scalability, and deployment, with lower overhead and better resource management.By selecting LLaMA models for our evaluation, we ensure that our benchmarks are representative of current state-of-the-art LLMs. LLaMA is one of the earliest open-source base models and is widely used in both industry and academia. Most models are fine-tuned based on LLaMA, making it an ideal choice for our evaluation model. Utilizing LLaMA allows our ELIB to support a broader variety of models, enhancing the versatility and applicability of our benchmarking framework.}

\begin{itemize}
\def\labelenumii{\alph{enumii}.}
\item[1]
  The LLM inference framework is employed for making predictions using a pre-trained transformer network. Various inference frameworks, including TensorRT \footnote{\url{https://developer.nvidia.com/tensorrt}}, ONNX-runtime\footnote{\url{https://github.com/onnx/tensorflow-onnx}}, and TensorFlow Lite\footnote{\url{https://www.tensorflow.org/lite}}, are utilized and undergo optimizations to enhance computational speed. Nevertheless, these frameworks are primarily designed for Nvidia GPU-based cloud inference servers, posing challenges for deployment on edge computing platforms due to requirements for high performance, portability, flexibility, and independence from third-party libraries. The open-source community is embracing such low-bit weight quantization and offers CPP-based implementations such as llama.cpp is based on GGML library\footnote{\url{https://ggml.ai}}. These implementations are typically optimized for CUDA and may not work on edge CPU and GPU. Therefore, it is important to address the challenge of making LLM inference efficient on an edge platform. However, as shown in Table 2, some representative lightweight inference frameworks are being embraced by the open-source community, making it feasible and elegant to deploy them on edge computing platforms.

\begin{itemize}
  \item[1]
   \emph{MLC-LLM}\footnote{\url{https://github.com/mlc-ai/mlc-llm}}: Machine learning compilation for LLM, serves as a universal deployment solution for enabling efficient operation of LLMs on edge devices by native hardware acceleration optimized by OpenCL, Vulkan, CUDA, and Metal. However, limitations in LLM model functionality like token streaming, only support quantization methods like LUT-GEMM  \cite{Park2022LGQ}, and a complex development process based on compiler like TVM present significant challenges when utilizing this approach.
  \item[2]
   \emph{CTranslate2}\footnote{\url{https://github.com/OpenNMT/CTranslate2}}: This is a C++ and Python library designed for efficient inference with Transformer models. It delivers fast and efficient execution on both CPU and GPU and supports multiple CPU architectures. The memory usage dynamically adjusts based on request size while maintaining performance standards through caching allocators on both CPU and GPU. However, it lacks support for adapters such as Lora, posing challenges in adapting new LLMs.
  \item[3] 
    \emph{Llama.cpp}\footnote{\url{https://github.com/ggerganov/llama.cpp}}: The primary objective of llama.cpp is to execute the LLaMA model using 4-bit integer quantization on edge computing platforms such as laptops and mobile phones. It is a pure C/C++ implementation without dependencies, optimized by OpenBLAS, OpenCL, CLblast, Accelerate, and Metal frameworks. The implementation supports mixed F16 and F32 precision, as well as 2-bit, 3-bit, 4-bit, 5-bit, 6-bit, and 8-bit integer quantization. Additionally, this framework was initially developed to efficiently infer LLaMA models. It is capable of loading and running models on a CPU, a key distinction from GPTQ \cite{Frantar2022GAP} models, which are loaded and run on a GPU. However, now some computing tasks of LLM can be offloaded to the GPU using llama.cpp to accelerate inference. This design approach positions llama.cpp as the optimal choice for inferring LLMs on edge computing platforms due to their efficiency, portability, flexibility, dependency, and complexity. In this work, we choose llama.cpp as the LLM inference frameworks.
  \end{itemize}

\item[2]
  LLMs are the dynamite behind the generative AI boom of 2023. While they have been in existence for some time, numerous open-source models such as LLaMA and closed-source models like GPT-3, have been introduced. When benchmarking LLMs on edge computing platforms, it is crucial to carefully consider an appropriate evaluation model that can accurately represent the key features of common LLMs.
  
  The LLaMA model family, introduced by Meta AI in February 2023, comprises four sizes (7B, 13B, 30B, and 65B). Since their release, LLaMA models have garnered significant attention from both research and industry communities. They have demonstrated outstanding performance on various open benchmarks, making them the most widely adopted open language models to date. Many researchers have expanded LLaMA models through instruction tuning or continual pretraining. Notably, instruction-tuned LLaMA models like Alpaca and Vicuna have emerged as a primary method for developing customized or specialized models, owing to their relatively low computational costs.

  Since the models are currently fully loaded into memory, we require enough disk space for storage and RAM for loading. Considering the size of the original and quantitation model as shown in Table 3, given that the RAM of our benchmarking device and most edge devices typically does not exceed 16GB, and the integer quantization model is smaller than the original model with a floating-point data type, the 7B parameters version is selected as the evaluation model in this work.
\end{itemize}

\begin{table}[!ht]
  \centering
  \caption{Storage of LLaMA models based on their parameters.}
  \begin{tabular}[]{ccc}
  \toprule
  Parameters & Original size & Quantized size (INT4) \\
  \midrule
  7B & 13 GB & 3.9 GB \\
  13B & 24 GB & 7.8 GB \\
  30B & 60 GB & 19.5 GB \\
  65B & 120 GB & 38.5 GB \\
  \bottomrule
  \end{tabular}
\end{table}

\subsection{Quantitation methods}

Quantization is a technique to reduce the numeric precision of neural network weights to lower the computation costs of inference. INT8 quantization \cite{Dettmers2022L8B} is the most widely-used approach today given the trade-off between high inference performance and reasonable model accuracy. However, outliers in activations have been observed and those outlier values are limiting the general adoption of INT8 quantization. FP8 \cite{Sun2019H8B} is a newly introduced data type that has attracted lots of attention while it has little adoption due to hardware unavailability. On the other hand, weight-only quantization becomes popular as it applies the low precision (e.g., 4-bit) to weights only, while keeping higher precision (e.g., 16-bit floating point) for activations, therefore maintaining the model accuracy. There are many excellent works on 4-bit weight-only quantization \cite{Banner2019PT4} that have demonstrated the effectiveness in LLM inference. Meanwhile, the open-source community is embracing such low-bit weight-only quantization and offers CPP-based implementations such as llama.cpp based on GGML library. These implementations are optimized by multiple backends like OpenBLAS, CUDA, Metal, OpenCL and achieve best performance by hardware and software co-optimization. As we choose llama.cpp as the inference framework for this work, we choose the corresponding quantitation method GGML for the best performance.

Table 4 summarizes the various types of quantization models by GGML. The naming convention for each type follows a specific format: qX\_Y, where X represents the number of bits used for weight storage (precision). A value of 0 for y indicates symmetric quantization without a zero point, while a value of 1 indicates asymmetric quantization, where each scale has a zero point. It is not recommended to use q\_2 or q\_3 version, as they significantly reduce model inference accuracy.

\begin{table}[!ht]
  \centering
  \caption{Common GGML quantization type}
\begin{tabular}[]{@{}
  >{\raggedright\arraybackslash}p{(\columnwidth - 2\tabcolsep) * \real{0.2}}
  >{\raggedright\arraybackslash}p{(\columnwidth - 2\tabcolsep) * \real{0.8}}@{}}
\toprule
Type & Use case \\
\midrule
q4\_0 & Original quant method, 4-bit \\
q4\_1 & Higher accuracy than q4\_0 but not as high as q5\_0.But has quicker inference than q5\_0 models\\
q5\_0 & Higher accuracy, higher resource usage, and slower inference \\
q5\_1 & Even higher accuracy, resource usage, and slower inference \\
q8\_0 & Almost indistinguishable from float16. High resource use and slow inference \\
\bottomrule
\end{tabular}
\end{table}

When examining the definition of q4\_0 in GGML quantization, weights are processed in blocks of 32 values. A scale factor (delta) is derived from the largest weight value within each block. Following this, all weights in the block are scaled, quantized, and efficiently packed for storage. This approach effectively reduces storage requirements and facilitates a relatively simple and predictable conversion between the original and quantized weights. This lightweight solution is suitable for performance optimization on edge computing devices.

\section{Benchmarking Methodology}

\begin{figure}[!ht]
  \centering
  \includegraphics[width=5.76806in,height=3.12431in]{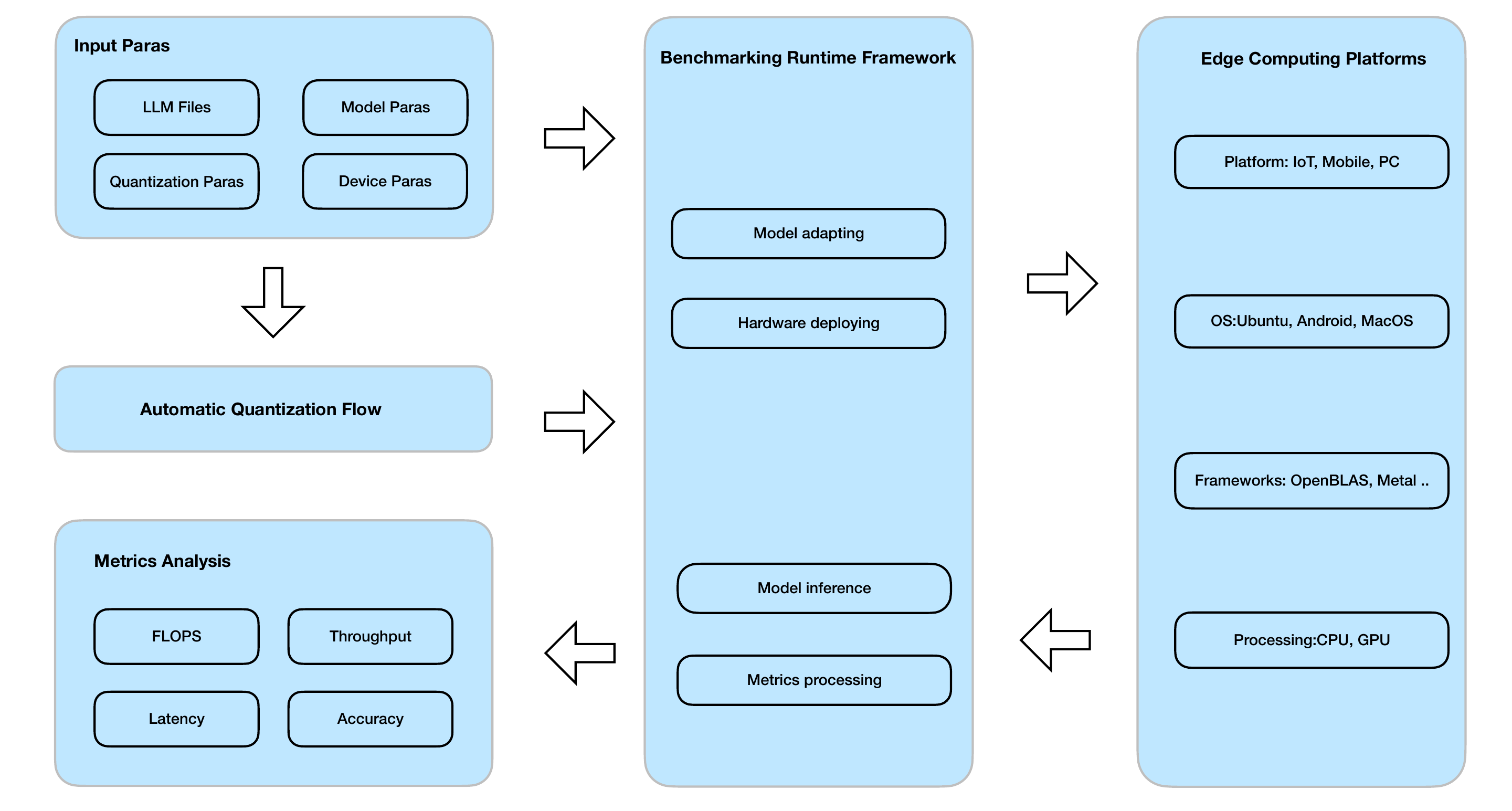}
  \caption{The design overview of ELIB indicates that the core component is the benchmarking runtime framework, including model adaptation, hardware deployment, model inference, and metrics processing}
\end{figure}

\begin{algorithm}
  \caption{Edge LLM Inference Benchmarking Program}
  \label{alg:benchmarking}
  \SetKwInput{KwInput}{Input}
  \SetKwInput{KwOutput}{Output}
  \SetKwProg{Try}{try}{:}{}
  \SetKwProg{Catch}{except}{:}{end}
  \KwInput{the original model file, quantization\_params, prompt\_data, benchmark\_params, device\_params}
  \KwOutput{performance metrics}
    config $\gets$ initialize() \;
    all\_quantized\_models $\gets$ automatic\_quantization\_flow(config.original\_model, config.quantization\_params) \;
    i $\gets$ 1 \tcp*{iteration number, starting from 1}
    \While{i $\leq$ config.benchmark\_params["iteration"] }{%
      \ForEach{quantized\_model \textbf{in} all\_quantized\_models}{%
        \tcp{model adaptation and hardware deployment}
        deployed\_model $\gets$ adapt\_and\_deploy\_model(quantized\_model, config.device\_params) \;
        \tcp{model inference}
        \Try{}{%
         inference\_result $\gets$ run\_inference(deployed\_model, config.test\_data, config.benchmark\_params) \;
        }
        \Catch{"time out" or "memory overflow" or "deadlock"}{%
          continue \;
        }
        flops $\gets$ calculate\_flops(inference\_result) \tcp*{float operations per second}
        throughput $\gets$ calculate\_throughput(inference\_result) \tcp*{tokens per second}
        latency $\gets$ calculate\_latency(inference\_result) \tcp*{milliseconds}
        accuracy $\gets$ calculate\_perplexity(inference\_result) \tcp*{perplexity score}
        mbu $\gets$ calculate\_mbu(inference\_result) \tcp*{mbu percentage}
      }
      \If(\tcp*[f]{all feasible models tested}){all\_quantized\_models $=\emptyset$}{
        break \;
      }
      i $\gets$ i + 1 \tcp*{proceed to the next iteration}
    }
\end{algorithm}

This section provides an overview of the design of ELIB, the benchmarking runtime framework, and the specific benchmarking metrics MBU that have been selected for optimizing evaluation.
Figure 1 provides an overview of ELIB. Initially, an original model with its associated benchmarking parameters (such as benchmarking configuration, model type, quantization requirements, hardware configuration, and etc.) is configured within ELIB. Subsequently, the quantization flow automatically quantizes the original model into different target quantized models. These target quantized models, along with their parameters, are then transferred to the benchmarking runtime framework of ELIB. This framework is responsible for model adaptation, hardware deployment, model inference, and processing benchmarking metrics. Following this, the computational task is transferred to the edge computing platform, and the performance data for multiple metrics is post-processed and analyzed by ELIB in a detailed manner to generate more readable and comprehensible benchmarking results.

\chen{
The top-level algorithm of ELIB is illustrated in Algorithm 1. The input parameters include the original model file (e.g., LLaMA 7B), quantization schemes (e.g., Q4\_0, Q5\_0, Q5\_1), test prompt data (e.g., user prompt files, Wikitext-2 data), benchmarking parameters (e.g., iteration, batch size, top-k, top-n, repeat\_last\_n), and device configurations (e.g., thread number, NEON, AVX, BLAS, GPU). The algorithm outputs performance metrics (e.g., flops, throughput, latency, and accuracy) that reflect the efficiency and accuracy of model inference under various conditions.
Given an original model, the algorithm first applies different quantization schemes to create a set of quantized models. These quantized models are then deployed and tested on the target edge device using the provided test data. The benchmarking parameters control the iterations, while the device parameters specify the hardware configuration.
}

\chen{
ELIB begins by initializing the configuration settings (Ln. 1) and proceeds to generate quantized models from the original model using the specified quantization schemes (Ln. 2). An iteration counter is initialized to manage the benchmarking loops (Ln. 3). For each iteration up to the specified limit (Ln. 4), the algorithm deploys each quantized model to the target device (Ln. 7) and performs inference using the provided test data (Ln. 10). Inference results are processed to compute various performance metrics(Ln. 13-17). Error handling mechanisms are in place to skip models that encounter runtime issues such as timeouts or memory overflows (Ln. 11-12).The algorithm checks if all quantized models have been tested. If so, it exits the loop (Ln. 18-19). Otherwise, it increments the iteration counter and continues the process (Ln. 20). 
}



\subsection{Benchmarking Runtime Framework}

The successful deployment of ELIB relies heavily on the benchmarking runtime framework, a lightweight edge-based LLM inference framework based on the core library of llama.cpp. This framework simplifies edge deployment with its straightforward and elegant structure. In contrast to the llama.cpp project, which is not very conducive to learning and secondary development, the benchmarking runtime framework has refactored many common components to enhance code readability and secondary development capabilities. This includes several practical methods, such as avoiding the consolidation of all logic and kernel code in a single file, separating framework code and kernel code to facilitate the optimization of kernel performance for specific edge platforms, etc.

\begin{figure}[!ht]
  \centering
  \includegraphics[width=5.76806in,height=1.04379in]{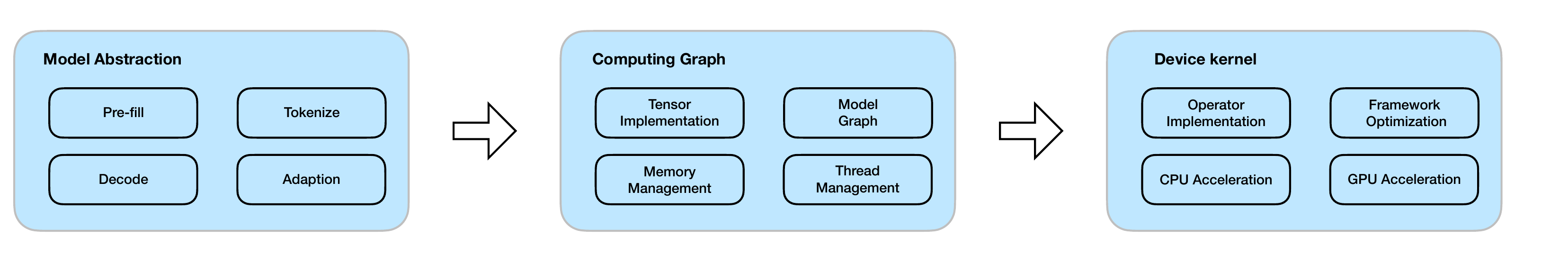}
  \caption{Model-Graph-Kernel structure of benchmark runtime framework offers a more elegant design methodology to deployment, operation, and expansion.}
\end{figure}

As Figure 2 shows the Model-Graph-Kernel structure of the benchmarking runtime framework. Model layer is used to store the input LLM parameters, tokenizer, historic tokens and etc. Graph consists of the implementation of certain LLMs, the abstraction of tensor library, basic algorithm operators, and the implementation and optimization of the KV cache shortage optimization system \cite{Shen2023ELI}. The kernel layer provides kernel computing code optimized for different edge platform backends, currently including OpenCL, Apple Metal, and naive CPU. When optimized kernels are not available, the system will directly fall back to running on the naive kernel.

The Model-Graph-Kernel design structure offers the benefits of easy deployment, efficient operation, and convenient expansion. It is easy to deploy, as the benchmarking runtime framework can be deployed without any dependencies. In terms of efficient operation, all computing tasks are performed natively on the edge device for optimal performance. Additionally, KV cache storage optimization creates an optimized KV cache with pre-allocated memory, updating only new tokens each time instead of loading all tokens. As for convenient expansion, the model layer can adapt more LLM types by parsing their model type, while the kernel layer can be ported to additional edge platforms by implementing the kernel function using platform-specific hardware APIs.

\subsection{Benchmarking Metrics}
Four commonly used benchmarking metrics and one novel metric are generated by ELIB to evaluate the inference performance. Different metrics are used for different aspects of the inference performance. FLOPS and throughput are used to evaluate compute - bound systems, while latency and MBU are used to optimize memory - bound systems. Accuracy is a key factor for optimizing other performance metrics.

\begin{enumerate}
\def\labelenumii{\alph{enumii}.}
\item
  \emph{FLOPS:} Floating-point Operations per Second is a metric used to assess the execution efficiency of edge platforms. By quantifying the number of floating-point operations executed per second, FLOPS enables the basic benchmarking of different platform efficiency across various large language model (LLM) implementations and edge platforms. 
\item
  \emph{Throughput:} The inference throughput of the LLM is crucial to the speed at which the model can generate output tokens, directly impacting user experience. This metric quantifies the time required to generate a response from the LLM model, calculated by dividing the total time taken by the number of runs and the number of tokens generated per run.
\item
  \emph{Latency:} The overall response latency can be measured using two metrics: TTLM and TTFT. TTLM (Time to load model) represents the time required for the device to load the complete model for the first user query, signifying the duration taken to load the large model from storage to RAM. This metric is typically influenced by hardware memory bandwidth and model size. TTFT (Time to First Token) indicates how quickly users begin to see the model's output after entering their query following the complete loading of the model. Minimal waiting times for a response are essential for real-time interactions. This metric is impacted by the time needed to process the prompt and generate the first output token.
\item
  \emph{Accuracy:} ELIB employs a common metric named perplexity score \cite{Huyen2019EMF}, which is the exponentiated average negative log-likelihood of a sequence and can be interpreted as an assessment of the model's ability to predict uniformly among a specified set of tokens in a corpus. A lower perplexity score indicates that the model is better at predicting the next word.
  \item
  \emph{MBU:} Inference of LLMs, particularly during the decoding process, is bottlenecked by the speed at which model parameters can be loaded from the device memory to the compute units. We introduce a novel metric known as Model Bandwidth Utilization (MBU). MBU values approaching $100\%$ suggest that the inference system is efficiently utilizing the available memory bandwidth. MBU serves as an improvement to the FLOPS utilization and throughput metrics for comparing different inference systems for both hardware and software in a normalized fashion, which holds great significance in memory-bound scenarios.
\end{enumerate}

 As shown in equation \ref{eq:mbu_def}, MBU is defined as the ratio of achieved memory bandwidth to peak memory bandwidth, theoretically indicating the most efficient utilization of the available memory bandwidth. Achieved memory bandwidth represents the actual bandwidth utilized by the LLM during inference, while peak memory bandwidth refers to the maximum bandwidth that the hardware can theoretically provide.

\begin{equation}
\text{MBU} = \frac{\text{Achieved Memory Bandwidth}}{\text{Peak Memory Bandwidth}}
\label{eq:mbu_def}
\end{equation}

As shown in equation \ref{eq:achieved_bandwidth}, achieved memory bandwidth is calculated as the sum of the total model parameter size and the KV cache size, divided by the total processing time. The total model parameter size refers to the overall size of the model's parameters, typically measured in bytes. The KV cache size represents the total size of the Key-Value cache, which stores intermediate results from the attention mechanism and scales with batch size, sequence length, and other model parameters. Time Per Output Token (TPOT) is the time taken to generate each output token during inference, measured in milliseconds per token. Typically, TPOT is the inverse of throughput.

\begin{equation}
\text{Achieved Memory Bandwidth} = \frac{\text{Total Model Parameter Size} + \text{KV Cache Size}}{\text{Time Per Output Token (TPOT)}}
\label{eq:achieved_bandwidth}
\end{equation}

The KV Cache, or Key-Value Cache, is a crucial component in transformer-based large language models (LLMs). During inference, each token in a sequence attends to all previous tokens, requiring the model to store intermediate results, specifically the keys and values for each token. By caching these key-value pairs, the model avoids recalculating them for each new token, significantly speeding up the inference process. The size of the KV cache can be calculated using equation \ref{eq:kv_cache}:

\begin{equation}
\text{KV Cache Size} = \text{Batch Size} \times \text{Sequence Length} \times \left(\frac{d_{\text{model}}}{n_{\text{heads}}}\right) \times n_{\text{layers}} \times n_{\text{kv\_heads}}  \times \text{Data Byte}  \times 2 
\label{eq:kv_cache}
\end{equation}

Batch Size refers to the number of sequences processed simultaneously during inference, while Sequence Length indicates the length of each input sequence in tokens. 
\( d_{\mathrm{model}} \) represents the dimension of the model's hidden states, and \( n_{\mathrm{heads}} \) refers to the number of attention heads in the model. 
\( n_{\mathrm{layers}} \) denotes the total number of layers in the model, and \( n_{\mathrm{kv\_heads}} \) specifies the number of heads used specifically for the KV cache. 
Data Byte refers to the number of bytes allocated for each data unit. The factor of 2 accounts for the storage of both keys and values in the cache.

\section{Experiment}

In this section, we utilize ELIB to benchmark three specific edge computing devices using five different quantization models with five benchmarking metrics, considering the various configurations discussed previously.
\subsection{Experiment Configurations}

As outlined in section 3, we have selected NanoPI, Xiaomi Redmi phone, and MacBook Air listed in Table 1 as the target benchmarking edge computing devices. Each device is accelerated by its corresponding frameworks (OpenBLAS, Acceleration, Metal, OpenCL, etc.). Furthermore, we will conduct separate benchmarking for the CPU model and the GPU hybrid model of the edge devices in order to fully harness the potential performance of the hardware.

\begin{table}[!ht]
  \caption{Quantized model for benchmarking}
\begin{tabular}[]{@{}
  >{\raggedright\arraybackslash}p{(\columnwidth - 8\tabcolsep) * \real{0.1285}}
  >{\raggedright\arraybackslash}p{(\columnwidth - 8\tabcolsep) * \real{0.1130}}
  >{\raggedright\arraybackslash}p{(\columnwidth - 8\tabcolsep) * \real{0.1291}}
  >{\raggedright\arraybackslash}p{(\columnwidth - 8\tabcolsep) * \real{0.1291}}
  >{\raggedright\arraybackslash}p{(\columnwidth - 8\tabcolsep) * \real{0.5002}}@{}}
\toprule
Quant method & Bits per weight & Model size & Max RAM required & Use Case \\
\midrule
q4\_0 & 4 & 3.5G & 6.1G & Original INT4 quant method \\
q4\_1 & 4.5 & 3.9G & 6.6G & Higher accuracy than q4\_0 but not as high as q5\_0. However, it has quicker inference than q5\_0 models \\
q5\_0 & 5 & 4.3G & 7.1G & Higher accuracy, higher resource usage and slower inference \\
q5\_1 & 5.5 & 4.7G & 7.6G & Even higher accuracy, resource usage and slower inference \\
q8\_0 & 8 & 6.7G & 8.6G & Almost indistinguishable from float16. High resource use and slow. Not recommended for most users \\
Original & 8 & 12.9G & 14.7G & Original model \\
\bottomrule
\end{tabular}
\end{table}

\begin{table}
  \centering
  \caption{Summary of experiment results: FLOPS measured by Giga for 4 threads and 8 threads are listed separately. Throughput is measured in token/s, latency is measured in seconds. MBU is the ratio of memory used by model, and accuracy is measured by perplexity score.}
  \tabcolsep=3pt
  \small
  \begin{tabular}{cccccccccccc}
    \toprule
     & \multirow{2}{*}{Platform} & \multirow{2}{*}{OS} & \multirow{2}{*}{Accelerator} & \multirow{2}{*}{Framework} & \multicolumn{2}{c}{FLOPS (Giga)} & Throughput & \multicolumn{2}{c}{Latency (Second)} & MBU & Accuracy \\ \cline{6-7} \cline{9-10}
     & & & & & t4 & t8 & (Tok/S) & TTLM & TTFT & (Ratio) & (Score) \\
    \midrule
    \multirow{9}{*}{Q4\_0} & \multirow{3}{*}{NanoPI}  & \multirow{3}{*}{Ubuntu}  & CPU & None            & 38.56   & 21.46   & 2.51  & 55.23  & 98.56  & 0.42 & 6.55  \\
                           &                          &                          & CPU & OpenBLAS        & 53.15   & 37.54   & 2.93  & 51.67  & 71.92  & 0.45 & 6.58  \\
                           &                          &                          & GPU & CLBlast\&OpenCL & 139.67  & 138.21  & 3.97  & 52.32  & 60.12  & 0.49 & 54.32 \\
                           & \multirow{3}{*}{Xiaomi}  & \multirow{3}{*}{Xiaomi}  & CPU & None            & 2.61    & 3.82    & 1.05  & 74.13  & 163.93 & 0.54 & 4.45  \\
                           &                          &                          & CPU & OpenBLAS        & 67.57   & 21.23   & 4.03  & 76.23  & 107.04 & 0.57 & 4.45  \\
                           &                          &                          & GPU & CLBlast\&OpenCL & 147.32  & 147.38  & 5.75  & 73.64  & 80.12  & 0.58 & 54.32 \\
                           & \multirow{3}{*}{Macbook} & \multirow{3}{*}{MacOS}   & CPU & None            & 443.62  & 401.49  & 8.21  & 7.21   & 20.72  & 0.62 & 6.58  \\
                           &                          &                          & CPU & Accelerate      & 676.60  & 466.47  & 14.63 & 6.82   & 12.86  & 0.71 & 6.58  \\
                           &                          &                          & GPU & Metal           & 1297.23 & 1291.29 & 19.72 & 7.28   & 8.10   & 0.82 & 6.58  \\
    \midrule
    \multirow{9}{*}{Q4\_1} & \multirow{3}{*}{NanoPI}  & \multirow{3}{*}{Ubuntu}  & CPU & None            & 38.11   & 21.33   & 1.67  & 67.12  & 101.65 & 0.59 & 6.56  \\
                           &                          &                          & CPU & OpenBLAS        & 55.29   & 38.36   & 2.71  & 62.06  & 74.50  & 0.61 & 6.32  \\
                           &                          &                          & GPU & CLBlast\&OpenCL & 137.22  & 133.58  & 3.47  & 65.28  & 56.23  & 0.65 & 59.12 \\
                           & \multirow{3}{*}{Xiaomi}  & \multirow{3}{*}{Android} & CPU & None            & 4.25    & 3.01    & 0.99  & 86.37  & 176.82 & 0.51 & 5.72  \\
                           &                          &                          & CPU & OpenBLAS        & 63.69   & 31.99   & 3.62  & 89.11  & 109.54 & 0.67 & 5.72  \\
                           &                          &                          & GPU & CLBlast\&OpenCL & 137.82  & 139.01  & 5.40  & 87.29  & 73.99  & 0.68 & 55.65 \\
                           & \multirow{3}{*}{Macbook} & \multirow{3}{*}{MacOS}   & CPU & None            & 337.76  & 321.77  & 7.92  & 7.89   & 28.45  & 0.71 & 6.44  \\
                           &                          &                          & CPU & Accelerate      & 722.59  & 456.66  & 13.46 & 7.03   & 14.13  & 0.79 & 6.44  \\
                           &                          &                          & GPU & Metal           & 1233.74 & 1234.22 & 18.70 & 7.67   & 10.35  & 0.85 & 6.43  \\
    \midrule
    \multirow{9}{*}{Q5\_0} & \multirow{3}{*}{NanoPI}  & \multirow{3}{*}{Ubuntu}  & CPU & None            & 37.91   & 20.55   & 1.56  & 81.16  & 115.23 & 0.51 & 6.55  \\
                           &                          &                          & CPU & OpenBLAS        & 58.69   & 38.36   & 2.67  & 77.11  & 82.81  & 0.54 & 7.20  \\
                           &                          &                          & GPU & CLBlast\&OpenCL & 135.22  & 139.28  & 2.64  & 81.93  & 49.12  & 0.58 & 61.33 \\
                           & \multirow{3}{*}{Xiaomi}  & \multirow{3}{*}{Android} & CPU & None            & 2.02    & 2.88    & 0.89  & 104.81 & 187.35 & 0.49 & 6.53  \\
                           &                          &                          & CPU & OpenBLAS        & 65.12   & 21.35   & 2.53  & 108.11 & 126.20 & 0.53 & 6.53  \\
                           &                          &                          & GPU & CLBlast\&OpenCL & 143.95  & 142.23  & 3.23  & 107.23 & 69.29  & 0.57 & 58.31 \\
                           & \multirow{3}{*}{Macbook} & \multirow{3}{*}{MacOS}   & CPU & None            & 407.21  & 387.25  & 4.23  & 8.89   & 31.49  & 0.67 & 6.27  \\
                           &                          &                          & CPU & Accelerate      & 753.71  & 461.36  & 11.14 & 8.12   & 14.44  & 0.71 & 6.27  \\
                           &                          &                          & GPU & Metal           & 1212.22 & 1214.72 & 18.31 & 8.34   & 12.34  & 0.84 & 6.43  \\
    \midrule
    \multirow{9}{*}{Q5\_1} & \multirow{3}{*}{NanoPI}  & \multirow{3}{*}{Ubuntu}  & CPU & None            & 37.51   & 20.11   & 1.45  & 85.23  & 129.33 & 0.51 & 6.45  \\
                           &                          &                          & CPU & OpenBLAS        & 57.92   & 38.95   & 2.31  & 83.62  & 84.46  & 0.54 & 6.29  \\
                           &                          &                          & GPU & CLBlast\&OpenCL & 133.19  & 132.87  & 2.63  & 89.11  & 45.61  & 0.55 & 63.44 \\
                           & \multirow{3}{*}{Xiaomi}  & \multirow{3}{*}{Android} & CPU & None            & 2.01    & 2.93    & 0.76  & 113.93 & 198.22 & 0.47 & 6.32  \\
                           &                          &                          & CPU & OpenBLAS        & 65.20   & 31.92   & 1.92  & 118.33 & 144.92 & 0.56 & 6.32  \\
                           &                          &                          & GPU & CLBlast\&OpenCL & 139.18  & 138.67  & 2.21  & 115.23 & 60.19  & 0.62 & 62.31 \\
                           & \multirow{3}{*}{Macbook} & \multirow{3}{*}{MacOS}   & CPU & None            & 422.62  & 410.89  & 7.23  & 10.42  & 37.33  & 0.71 & 6.33  \\
                           &                          &                          & CPU & Accelerate      & 750.89  & 459.48  & 10.19 & 9.28   & 15.55  & 0.79 & 6.33  \\
                           &                          &                          & GPU & Metal           & 1187.23 & 1188.33 & 16.11 & 9.39   & 13.94  & 0.85 & 6.42  \\
    \midrule
    \multirow{9}{*}{Q8\_0} & \multirow{3}{*}{NanoPI}  & \multirow{3}{*}{Ubuntu}  & CPU & None            & 37.16   & 19.25   & 1.39  & 93.87  & 139.65 & 0.53 & 6.90  \\
                           &                          &                          & CPU & OpenBLAS        & 54.56   & 39.39   & 2.12  & 89.66  & 88.80  & 0.57 & 6.27  \\
                           &                          &                          & GPU & CLBlast\&OpenCL & 130.87  & 129.38  & 2.42  & 88.57  & 50.39  & 0.61 & 67.56 \\
                           & \multirow{3}{*}{Xiaomi}  & \multirow{3}{*}{Android} & CPU & None            & 3.76    & 4.77    & 0.67  & 121.65 & 210.82 & 0.48 & 6.27  \\
                           &                          &                          & CPU & OpenBLAS        & 66.15   & 31.61   & 1.80  & 128.34 & 159.56 & 0.55 & 6.27  \\
                           &                          &                          & GPU & CLBlast\&OpenCL & 138.11  & 137.37  & 2.00  & 127.24 & 54.38  & 0.57 & 68.12 \\
                           & \multirow{3}{*}{Macbook} & \multirow{3}{*}{MacOS}   & CPU & None            & 442.63  & 387.29  & 4.23  & 12.23  & 40.12  & 0.63 & 6.31  \\
                           &                          &                          & CPU & Accelerate      & 759.46  & 455.03  & 8.60  & 11.22  & 17.40  & 0.72 & 6.31  \\
                           &                          &                          & GPU & Metal           & 1180.64 & 1181.17 & 15.89 & 11.39  & 16.32  & 0.87 & 6.42  \\
    \bottomrule
  \end{tabular}
\end{table}

\subsubsection{Quantitation Models}

As mentioned in section 3, the benchmarking model selected is the LLaMA 7B parameters version listed in Table 3, which is 13GB in size and uses the FP16 data type. We applied GGML to quantize the LLaMA 7B using five commonly used quantization methods with different bits per weight for the target benchmarking. The quantized models resulted in variations in file size and bits per weight. It is important to highlight that the selection of the quantization method has a significant impact on the performance and accuracy of the model.
Table 5 displays the file size and the associated bit weight of each quantized model, illustrating that larger model file sizes typically result in improved accuracy but slower inference speed. For edge devices with limited performance capabilities, q4\_0 is recommended due to its favorable balance between accuracy and inference speed. Conversely, for edge devices with higher performance capabilities, q8\_0 is advised, as it has the potential to maintain satisfactory accuracy while offering improved inference speed.

\subsection{Results and Analysis}

All benchmark results are listed in Table 6. And we conduct extensive comparisons and visualize the data in figures to facilitate detailed discussions and provide the practical prospective heuristics.

As explained earlier, the inference of LLM is bottlenecked by memory bandwidth. The goal of optimizing the inference performance is to improve MBU with the help of other metrics without sacrificing accuracy. To evaluate MBU, we consider three research questions:
\begin{itemize}
    \item \textbf{RQ1:} What is the key factor in maximizing MBU?
    \item \textbf{RQ2:} What is the key constraint in maximizing MBU?
    \item \textbf{RQ3:} What is the key unpredictability in maximizing MBU?
\end{itemize}

\begin{figure}[!ht]
  \centering
  \subfigure[]{\includegraphics[width=3.2in,height=2.1in]{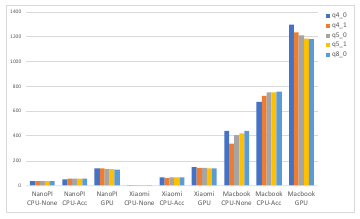}}
  \subfigure[]{\includegraphics[width=2.5in,height=2.1in]{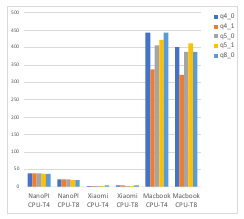}}
  \caption{(a) The comparison of FLOPS measured in billions of floating-point operations per second (GFLOPS), is presented between the non-accelerated and accelerated versions for three platforms across five testing quantization models. (b) Provides a comparison of the FLOPS measured in GFLOPS for 4 threads and 8 threads.}
\end{figure}

\begin{enumerate}
  \item
 
    \emph{FLOPS Performance:} The computations in LLMs are primarily characterized by matrix multiplication operations, which are typically constrained by the number of computing units and memory bandwidth on edge hardware. Therefore, we aim to utilize matrix-matrix multiplication to evaluate the general FLOPS performance of a specific edge device.
    There is substantial variation in the overall FLOPS performance across different hardware platforms and acceleration frameworks. Figure 3a indicates the accelerated frameworks like OpenBLAS and Apple Accelerate demonstrate notable performance improvements when compared to non-accelerated CPU FLOPS, especially for the Android platform and large bit-width models. This is attributed to the increased computational workload, enabling a higher number of parallel operations. Moreover, the performance of GPU in terms of FLOPS is superior to that of CPU across all platforms and data types. Consequently, more computing cores can engage in simultaneous processing, leading to the concurrent execution of a greater number of operations.Using acceleration libraries such as OpenBLAS can significantly enhance the FLOPS performance of the CPU, and hybrid computing like GPU can improve overall computing power. However, the configuration and implementations of acceleration libraries may vary slightly among different vendors, it's important to test vendor implementations before using them.

    From Figure 3b we find the performance of using four threads is observed to slightly outperform that of using eight threads. This counterintuitive finding contradicts the expectation that greater CPU thread utilization should yield higher FLOPS performance. The reason behind this is attributed to suboptimal parallel computing optimization and memory bandwidth.Optimizing parallelization and memory usage is crucial for hardware-level optimization. Understanding low-level hardware details is essential for identifying the causal performance bottlenecks.

    \chen{
    For RQ1, we observe that computations are predominantly dominated by matrix-matrix multiplication operations, which are typically memory-bandwidth-bound on most hardware instead of computing-peak-bound. Consequently, the speed of token generation depends more on how quickly model parameters can be loaded from memory to local caches/registers rather than on the computation speed. Therefore, both in theoretical calculations and experimental data, a higher MBU value consistently correlates with higher FLOPS and throughput. To achieve the maximum theoretical MBU, three key strategies should be employed. The first is increasing the batch size. Larger batch sizes allow more data to be processed simultaneously, which increases the achieved memory bandwidth. The second strategy is optimizing the sequence length. By adjusting the sequence length to an optimal value, it is possible to balance computation load and memory usage, thereby enhancing memory bandwidth utilization. The third strategy involves efficient KV cache management. Minimizing the KV cache size through quantization or optimized cache strategies can reduce the memory footprint, freeing up more memory bandwidth for other operations.
    }

\begin{figure}[!ht]
  \centering
  \includegraphics[width=2.5in,height=2.1in]{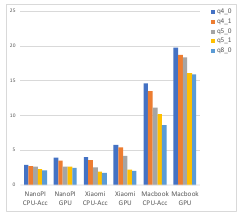}
  \caption{Inference throughputs results, measured in tokens per second.}
\end{figure}

\item
    \emph{Inference Throughputs:} Figure 4 illustrates the throughput results obtained from benchmarking. All devices demonstrate improved throughput performance with low-bit quantized models. For instance, on the NanoPI, the token throughput of q4\_0 was 1.38X and 1.64X higher than q8\_0 for accelerated CPU and GPU, respectively. On the Xiaomi, the ratio was 2.23X and 2.88X, and on the MacBook Air, the ratio was 1.7X and 1.24X. However, low-bit quantized models with better inference throughputs exhibit poor accuracy.

    By utilizing GPU acceleration to expedite model inference, all devices demonstrate enhanced throughput performance. Specifically, the average GPU throughput for the Nano Pi is 1.18X higher than that of the CPU, while for the Xiaomi, the ratio is 1.41X, and for the MacBook Air, the ratio is 1.53X. Furthermore, not only have significant throughput improvements been achieved, but some of the workload has also been effectively offloaded from the CPU to the GPU. Hybrid computing allows the CPU to be freed for other computing tasks to enhance user experience. However, in some cases, due to inconsistencies in data precision between the GPU and CPU across different hardware configurations and vendor implementations, the accuracy of GPU-accelerated inference may not be as good as that of CPU inference.Model quantization and hybrid computing can substantially enhance inference throughputs at the expense of accuracy. It is important to make an appropriate trade-off between throughput and accuracy for specific applications.
    
\begin{figure}[!ht]
  \centering
  \subfigure[]{\includegraphics[width=3.2in,height=2.1in]{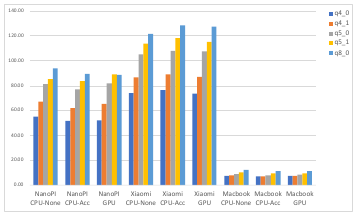}}
  \subfigure[]{\includegraphics[width=3.2in,height=2.1in]{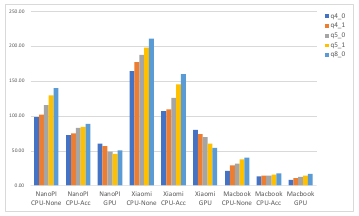}}
  \caption{(a) The time measured in second required to load a model (TTLM) on each device varies for different quantized models. (b) The time it takes to the first token (TTFT) after the user input, measured in second.}
\end{figure}

\begin{figure}[!ht]
  \centering
  \includegraphics[width=3.1in,height=2.1in]{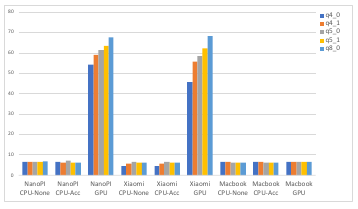}
  \caption{Inference accuracy measure in perplexity score.}
\end{figure}

\item
    \emph{Latency: }Figure 5a indicates the time to load model (TTLM) for each device and quantized model. In the case of a specific model, a device with greater RAM bandwidth, such as the MacBook Air, loads the model more rapidly than a device with smaller RAM bandwidth, such as the NanoPI and Xiaomi. This is particularly beneficial for certain applications, such as chatbots, where low latency for rapid responses is a top priority in order to provide an enhanced user experience. Additionally, as shown in Figure 5b, time to first token (TTFT) is affected by throughput and input length. Devices with high throughput to inference a low-bit model typically have a smaller TTFT. The overall latency exhibits a positive correlation with the model size, device bandwidth, MBU, and desired input.
    
    The factor of data batch size is crucial in optimizing the tradeoff between throughput and latency. From sections 5.2.2 and 5.2.3, we know that increasing the data batch size can maximize throughput, resulting in a 14x increase, but also causes a 4x increase in latency. Shared inference services on edge computing platforms typically select a balanced batch size. However, after a certain batch size, when we enter the compute-bound area, increasing the batch size further increases latency without increasing throughput.

    \chen{
    For RQ2, when optimizing MBU to improve throughput and latency, two key constraints must be considered to ensure that the system's performance and memory usage remain within acceptable limits. The first constraint is memory capacity: the combined size of model parameters and the KV cache should not exceed the hardware's available memory. Exceeding this limit can result in memory overflow or inefficient resource utilization, leading to potential system bottlenecks. The second constraint is total latency, which is the sum of the time taken to generate the first token (TTFT) and the time required to generate each subsequent token multiplied by the total number of output tokens (N). This total latency must not exceed the required latency to ensure the system meets performance expectations, particularly in real-time applications. This ensures that the system's overall latency meets the specified targets, which is critical for real-time applications.
    }
\item
    \emph{Accuracy:} As Figure 6 shows, after 100 inference iterations, all devices achieved nearly identical perplexity scores between 4-8 for the CPU model, including both the non-accelerated and accelerated versions, indicating high accuracy. The perplexity scores of the Metal-supported GPU-accelerated version for MacBook are nearly identical to those of the CPU version, suggesting equivalent inference accuracy. However, when utilizing the OpenCL-supported GPU on the Xiaomi phone and NanoPI, the perplexity score was almost 10 times higher compared to the CPU inference, indicating significantly poorer inference accuracy. Additionally, larger bit-width models such as q8\_0 resulted in higher perplexity scores for the GPU implementations using OpenCL. The decline in accuracy was likely attributed to suboptimal parallelization design and data precision issues when using OpenCL on these configurations.Hardware configurations like GPU acceleration for model inference needs to be rigorously tested on each platform and framework combination. There are substantial differences between configurations that can impact data precision and hybrid computing result.

    \chen{
    For RQ3, while enhancing MBU and applying quantization can improve performance and reduce latency, accuracy is expected to remain predictable, similar to other metrics such as FLOPS, throughput, and latency. However, on edge devices, a more unpredictable and potentially catastrophic issue for accuracy is the level of GPU hardware support and software configuration on edge computing devices. Given the numerous edge computing hardware platforms and the significant differences in software and hardware implementations, it is challenging to describe and test accuracy with a single factor. The most effective engineering approach remains comprehensive, integrated evaluation and optimization.
    }
\end{enumerate}

\subsection{\chen{Threats to Validity}}

\chen{
The primary threat to internal validity is model performance evaluation. The performance of large models on edge devices can be influenced by various factors, such as hardware specifications, software optimizations, and specific configurations of the deployment environment. To mitigate this threat, we conducted extensive benchmarking across multiple devices and configurations, ensuring that our results are representative of a wide range of scenarios.}

\chen{
External validity is challenged by the generalizability of our findings. The findings from our experiments on specific edge devices may not be directly applicable to all types of edge hardware. To address this issue, we selected a diverse set of edge devices, including low-end, mid-range, and high-end categories, and verified our results across these different categories to enhance the generalizability of our conclusions.}

\chen{
Construct validity concerns the definition of success metrics. The choice of success metrics (e.g., latency, throughput) might not capture all aspects of model deployment performance. To address this, we included additional factors and discussed their relevance to various deployment scenarios, ensuring a more comprehensive evaluation.}

\chen{
Finally, conclusion validity is reliant on replication. The ability to replicate our results is crucial for validation. In our open-source project, we provided detailed documentation of our experimental setup, including hardware and software configurations, datasets used, and step-by-step procedures. This comprehensive documentation enables other researchers to replicate our study, thereby enhancing the reliability of our findings.
}

\section{Conclusion}

In order to address the challenges associated with evaluating LLMs on edge computing platforms, we introduce a benchmarking methodology and a tool called ELIB. ELIB integrates multiple acceleration frameworks, an automatic quantization workflow, and a benchmarking runtime framework. We utilize ELIB to benchmark three representative edge computing platforms using five quantization models, and then analyze the benchmarking results to obtain practical heuristics that can guide future work on deploying LLMs on edge computing platforms. From this work we know that the optimization goal of inferring different quantized LLMs on multiple edge computing platforms is driven by application-specific requirements. Balancing interactive performance, maximizing throughput, minimizing latency, and high accuracy should be carefully considered to achieve the best inference performance through the tuning of the combination of quantized LLMs, acceleration frameworks, and edge computing hardware. In the future, we plan to expand ELIB to support a wider range of LLM types, quantization methods, and edge computing platforms, in order to provide more practical tools.





\bibliographystyle{elsarticle-num} 
\bibliography{refs}







\end{document}